\documentclass{PoS}

\title{Spectroscopy of mesons with bottom quarks}

\ShortTitle{Spectroscopy of mesons with bottom quarks}

\author{\speaker{Sin\'ead M. Ryan}\\
  School of Mathematics and Hamilton Mathematics Institute,
  Trinity College, Dublin 2, Ireland\\
        E-mail: \email{ryan@maths.tcd.ie}}

\abstract{Preliminary results for the spectra of excited and
  exotic bottom mesons are presented. The calculation on a dynamical
  anisotropic ensemble exploits distillation, enabling the use of a large
  basis of
  interpolating operators including those proportional to the gluonic
  field strength which are relevant for hybrid states. A comparison of results
  with similar calculations in the light and charm sectors is discussed.}

\FullConference{37th International Symposium on Lattice Field Theory -
  Lattice2019\\
  16-22 June 2019\\
  Wuhan, China}

\begin{document}

\section{Introduction}
The heavy quark spectrum has proved a fertile hunting ground for
exotic strongly-bound states. 
The discoveries, beginning more than fifteen years ago, of unexpected and
surprisingly narrow charmonium-like states close to and above strong decay thresholds
has spurred theoretical investigations into the nature and structure of these so-called XYZ states
which confront quark models and motivate model-independent calculations of QCD to fully understand the nature of the
strong force responsible for binding in hadrons. 
The phenomenology and experimental properties of these states are described in Ref~\cite{Lebed:2016hpi},
while a recent summary of lattice spectroscopy and scattering in charmonium and
bottomonium is in Ref.~\cite{Padmanath:2019wid} and references therein. While a
resolution of the puzzles in the charmonium sector still eludes us, further
puzzles appear in the bottomonium sector. To date five hadronic states with $b\bar{b}$ content have
been discovered which are inconsistent with a simple $q\bar{q}$ description. While three have conventional
quantum numbers - $\Upsilon(10580, 10860, 11020)$ - a further two, $Z_b(10610, 10650)$, appear to require four
quarks, possibly in a tetraquark or molecular arrangement. The bottomonium meson spectrum remains relatively poorly
mapped out by experiments although LHCb and Belle II are expected to address this~\cite{Kou:2018nap,Bediaga:2018lhg}.
Meanwhile, lattice calculations
provide some information but a complete picture of the spectrum
(even for the $b\bar{b}$-like states) is missing. In Refs.~\cite{Liu:2012ze,Cheung:2016bym} we performed an
extensive calculation of the excited and exotic charmonium spectrum, determining all states up to
$J=4$ including a supermultiplet of hybrid states. This motivates the exploratory study of the
excited and exotic bottomonium spectrum presented here.

\section{Calculation details}
The gauge action
is Symanzik-improved (anisotropic) with tree-level tadpole improved
coefficients and $N_f=2+1$. The sea and valence quarks, including the bottom quark, are
simulated with an anisotropic clover action with stout-smeared~\cite{Morningstar:2003gk} spatial links.
Details of the tuning of the action parameters are given in
Refs.~\cite{Edwards:2008ja, Lin:2008pr} while the mass-dependent tuning of the
valence anisotropy, for a simulation with charm quarks, is described in
Ref.~\cite{Liu:2012ze}. In that latter study, the relativistic anisotropic action was successfully
used to determine the excited and exotic charmonium
spectrum while in Ref.~\cite{Moir:2013ub} a similar calculation yielded the excited open-charm
spectrum and Ref.~\cite{Cheung:2016bym} included an analysis of the light quark mass dependence.
This study includes a mass-dependent tuning of the anisotropy in the bottomonium sector and
distillation~\cite{Peardon:2009gh} to facilitate the construction of large operator bases
in the relevant lattice irreducible
representations (irreps, $\Lambda^{PC}$). Using the variational procedure the energies of
states in each $\Lambda^{PC}$ are determined and exploiting a spin assignment method
informed by the operator overlaps
the spectrum is determined up to continuum spin $J=4$, including hybrids and states with
manifestly exotic quantum numbers. Further details of the operator construction and spin
identification are given in Ref.~\cite{Liu:2012ze} and earlier work by the Hadron Spectrum
Collaboration. Some relevant parameters and action details are given in Table~\ref{tab:details}.
\begin{table}[h]
  \begin{center}
  \begin{tabular}{ccccccc}
    Volume & $M_\pi$ (MeV) & $\xi$ & $a_s$ & $a_t^{-1}(m_\Omega)$ & $N_{\rm configs.}$ & $N_{\rm vecs}$ \\
    \hline
    $20^3\times 128$ & 391 & 3.5 & $\sim$0.12fm & $\sim$5.67 GeV & 603 & 128
  \end{tabular}
  \caption{Details of the ensemble used in this study. The anisotropic lattice volume corresponds to
    $(L/a_s)^3\times(T/a_t)$ while $N_{\rm configs.}$ and $N_{\rm vecs}$ refer to the number of gauge field
    configurations and the number of distillation eigenvectors respectively.
    This ensemble has been used in previous work by the
    Hadron Spectrum Collaboration,
    including Refs.~\cite{OHara:2017eoq,Dudek:2013yja}, where further details can be found.}
  \label{tab:details}
  \end{center}
\end{table}
\subsection{Dispersion relations}
On the anisotropic lattice $a_tm_b < 1$, however $a_sm_b > 1$
which could in principle lead to large discretisation errors. The effect of such artefacts is
investigated by determining the dispersion relations
of heavy-heavy (bottomonium) and heavy-light (B meson) systems, from which 
parameters including the measured anisotropy and the meson masses (rest and kinetic) are determined. 

\begin{figure}[h]
  \begin{center}
  \includegraphics[width=0.65\textwidth]{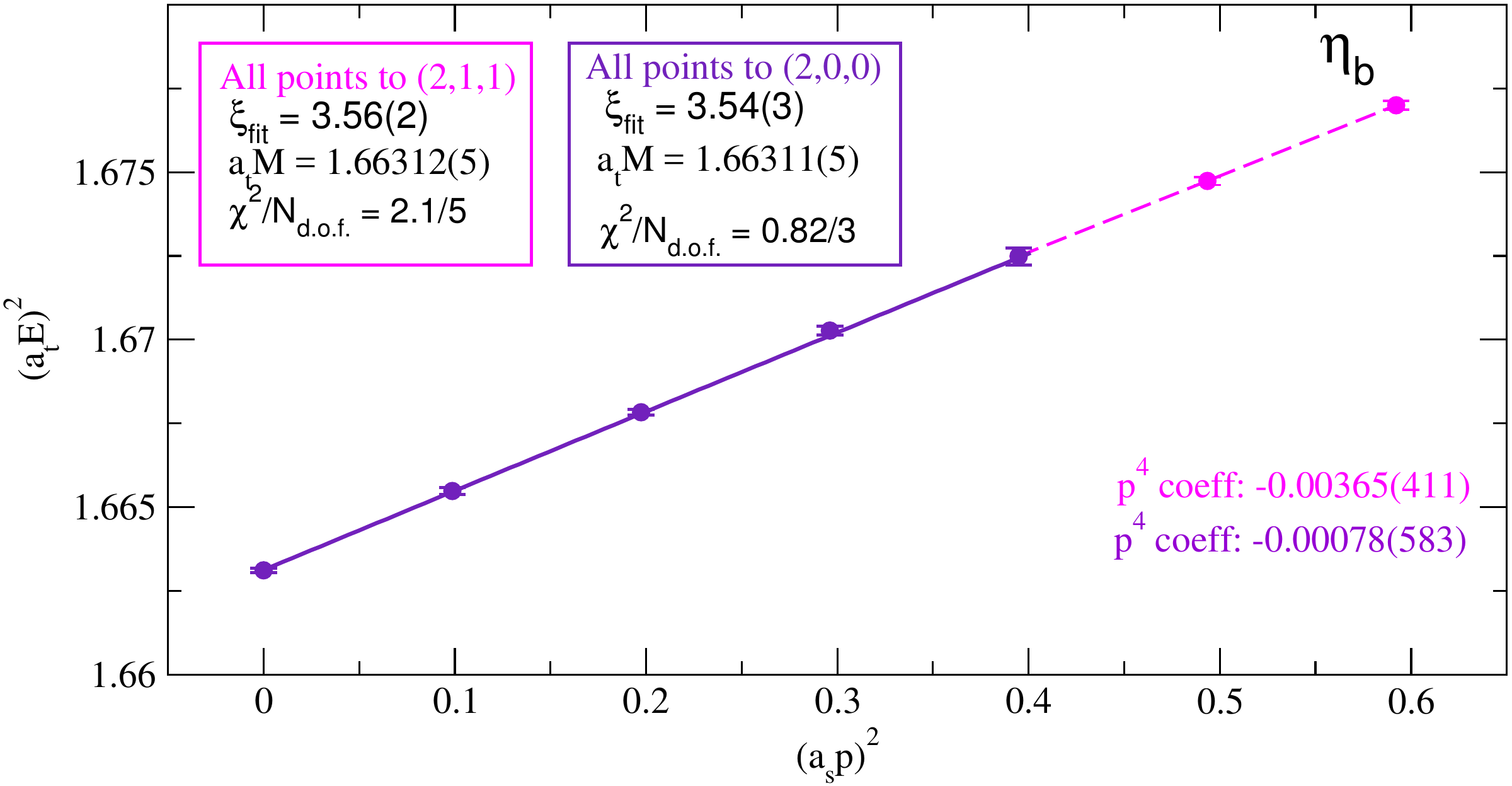}
  \caption{The dispersion relation for the $\eta_b$ with the input heavy
    quark mass tuned so that $m_{\eta_b}$ determined from the intercept takes
    its experimental value and the input anisotropy yields a measured value
    close to the target value of 3.5. The tuned values were taken from fits to
    points up to $(2,0,0)$ units of momenta (purple) although little change was seen
    when higher momenta (magenta) were included, as indicated. The errors are statistical only.}
  \label{fig:etab_disp}
  \end{center}
\end{figure}

\begin{figure}[h]
  \begin{center}
  \includegraphics[width=0.45\textwidth]{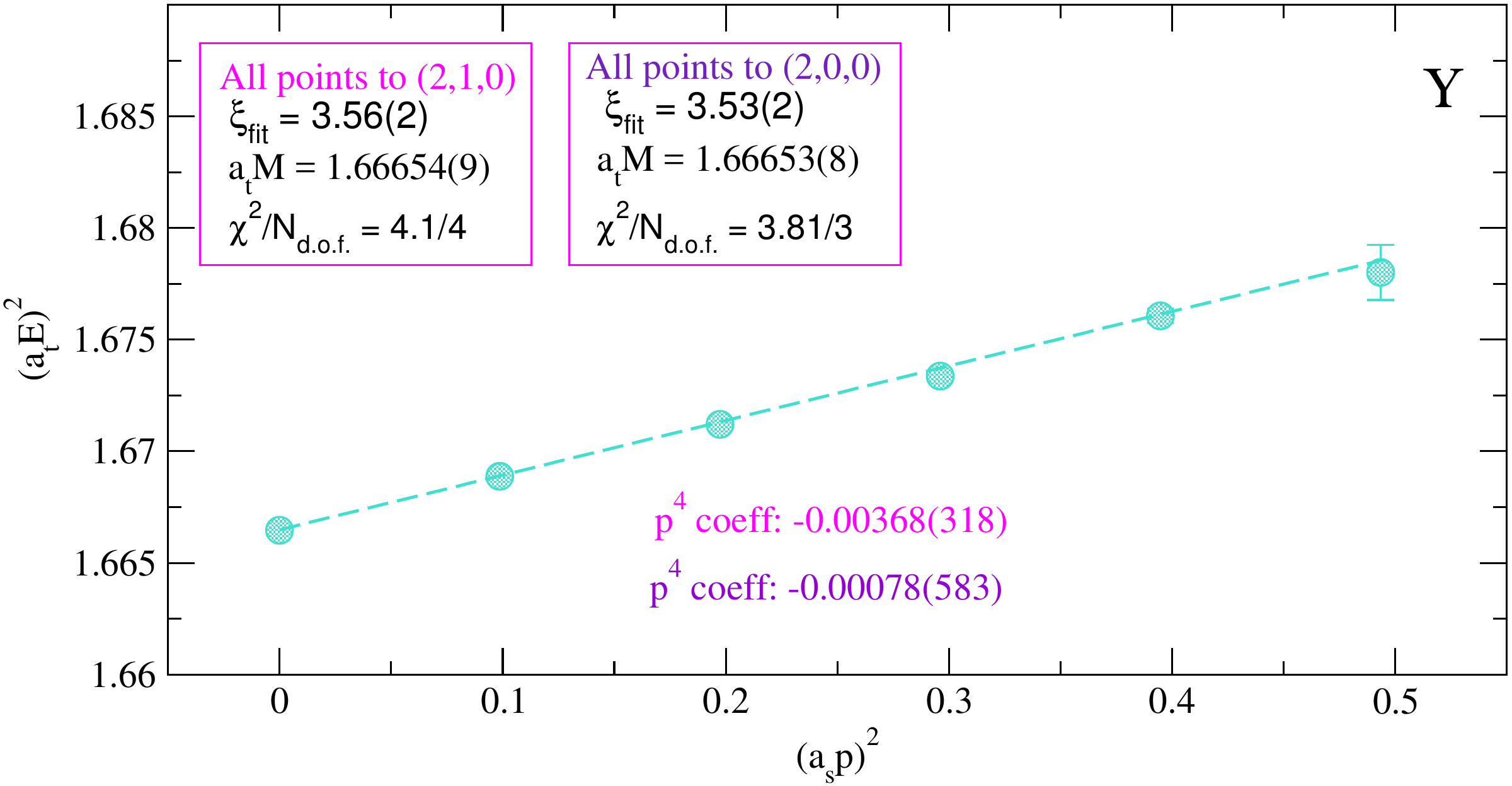}
  \includegraphics[width=0.45\textwidth]{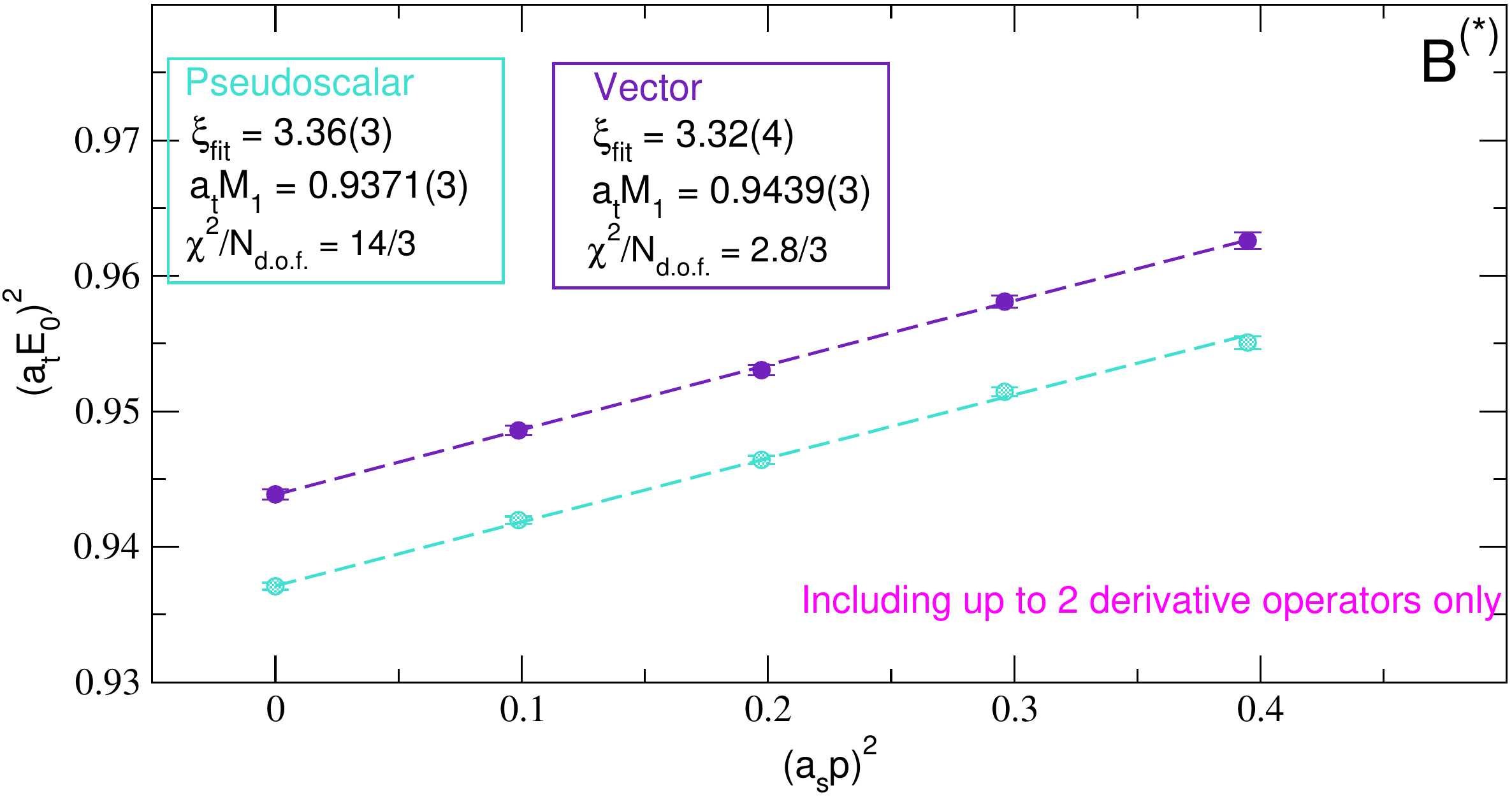}
  \caption{Dispersion relations for the $\Upsilon$ (left pane) and the
    heavy-light pseudoscalar and vector mesons (right pane). The
    bare heavy quark mass and the valence anisotropy have been determined
    from the $\eta_b$ dispersion relation. The data show no deviation from
    linear (relativistic) behaviour and the rest and kinetic masses determined
    in each case are in agreement. The anisotropy measured from the slope of the
    dispersion relations (shown in the boxes in each plot) is also in good agreement with the result for the
    $\eta_b$ meson.} 
  \label{fig:other_disp}
  \end{center}
\end{figure}

The relativistic dispersion relation of a meson on an anisotropic lattice with quantised momenta is
\begin{equation}
  \left(a_tE\right)^2 =\left(a_tM\right)^2 + \left(\frac{1}{\xi}\right)^2(a_sp)^2,
\end{equation}
where $a_s$ and $a_t$ are the spatial and temporal lattice spacings respectively and momenta are quantised on 
a lattice of length $L$ with periodic boundary conditions according to
$a_s\vec{p}=\frac{2\pi}{L}\left(n_x,n_y,n_z\right)$, with $n_i\in\{0,1,\ldots ,L/a_s -1\}$. 
In this study the heavy quark mass and the valence (heavy) anisotropy are tuned so that the dispersion
relation of the $\eta_b$ meson, including momenta up to $(2,0,0)$,
reproduces the experimentally measured $\eta_b$ mass and the target
anisotropy $\xi=3.5$. This is shown in Figure~\ref{fig:etab_disp}, which
also shows the fit parameters determined
when momenta up to $(2,1,1)$ are subsequently included in a separate fit and
when a term proportional to $p^4$ is allowed in the fits. The effect of these
additional points and terms is neglible and fits yield consistent relativistic dispersion relations. 
Taking the bare heavy quark mass and the tuned anisotropy from the $\eta_b$ dispersion relation
the $\Upsilon$ and the heavy-light pseudoscalar and
vector dispersion relations are determined, shown in Figure~\ref{fig:other_disp}.
The measured anisotropies are consistent and
in agreement with the target value, resulting in consistent rest and kinetic
meson masses. Once again, higher-order terms in the lattice dispersion relation,
${\cal O}(p^4)$, were included
in the fits and the effects were found to be negligible. 
\section{Results}
\begin{figure}[h]
  \begin{center}
    \includegraphics[width=0.22\textwidth]{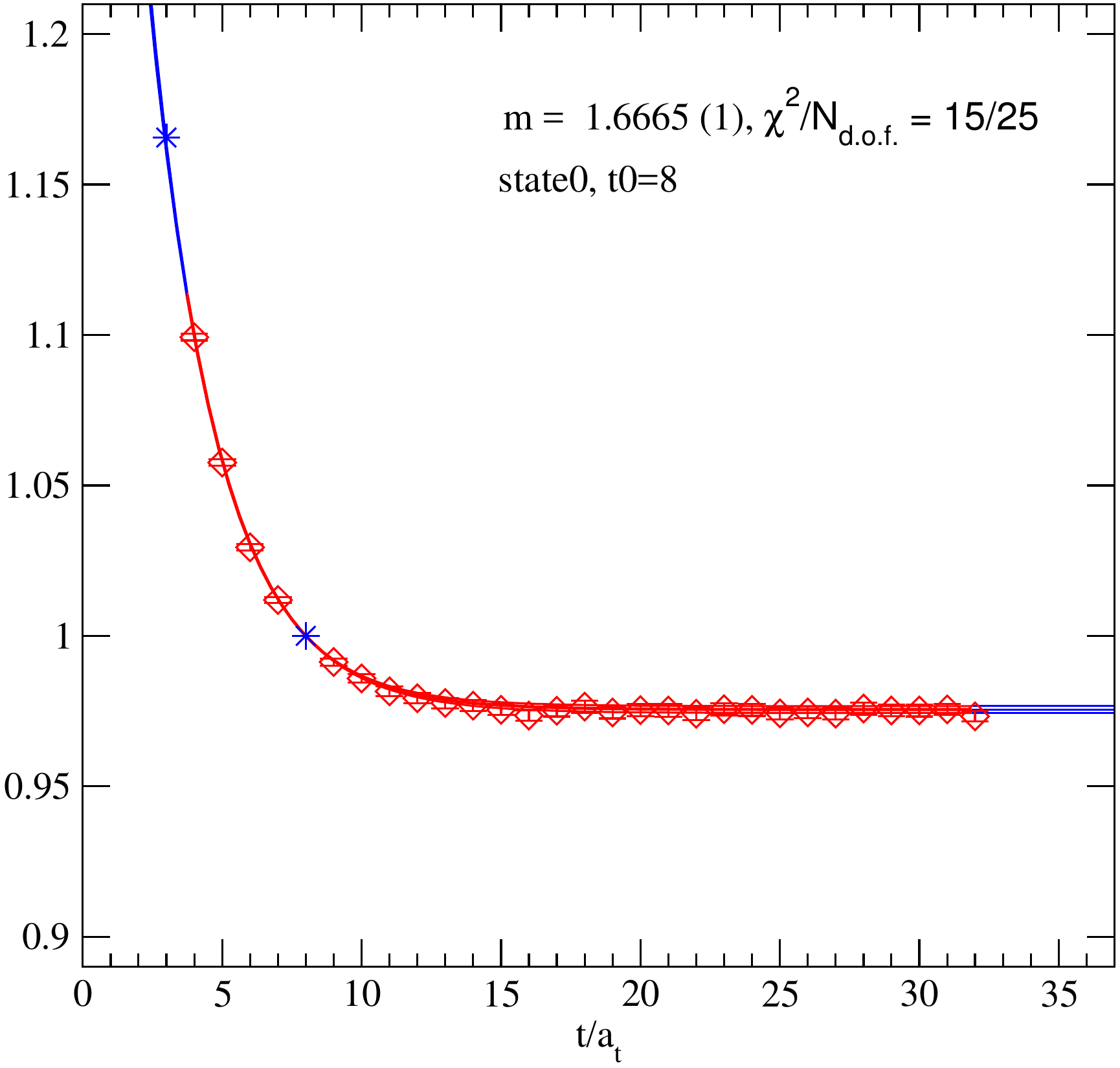}
    \includegraphics[width=0.22\textwidth]{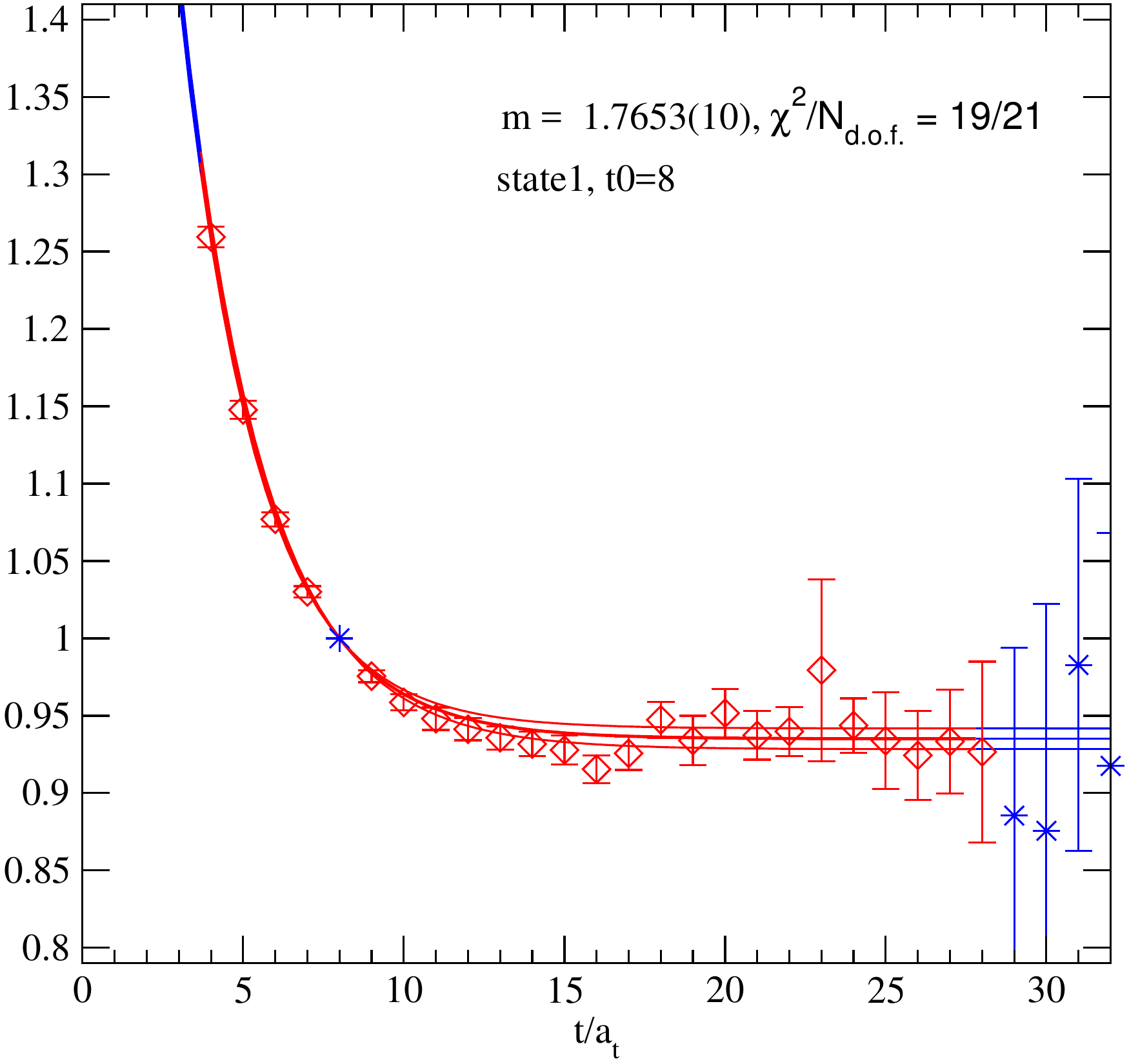}
    \includegraphics[width=0.22\textwidth]{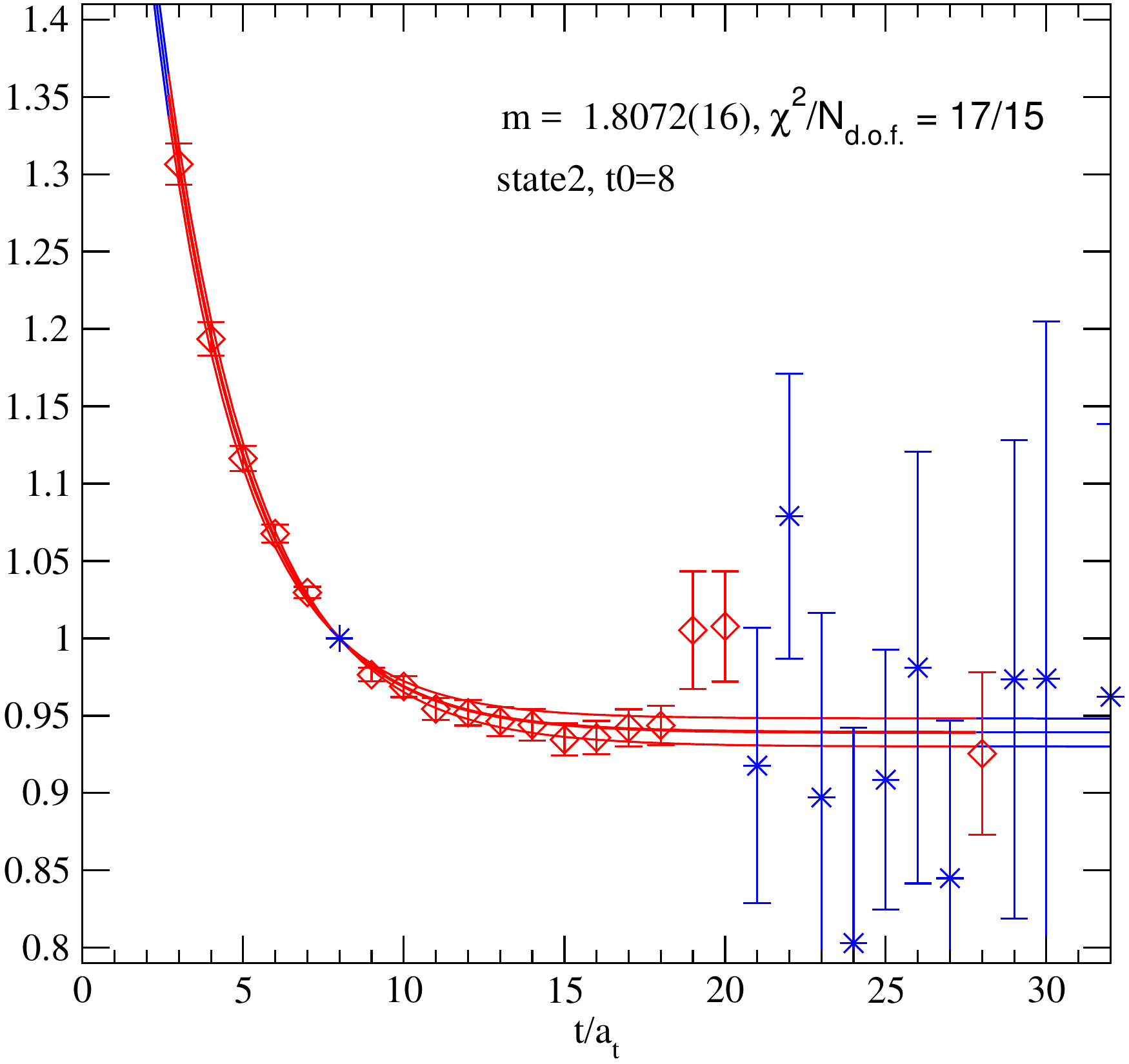}
    \includegraphics[width=0.22\textwidth]{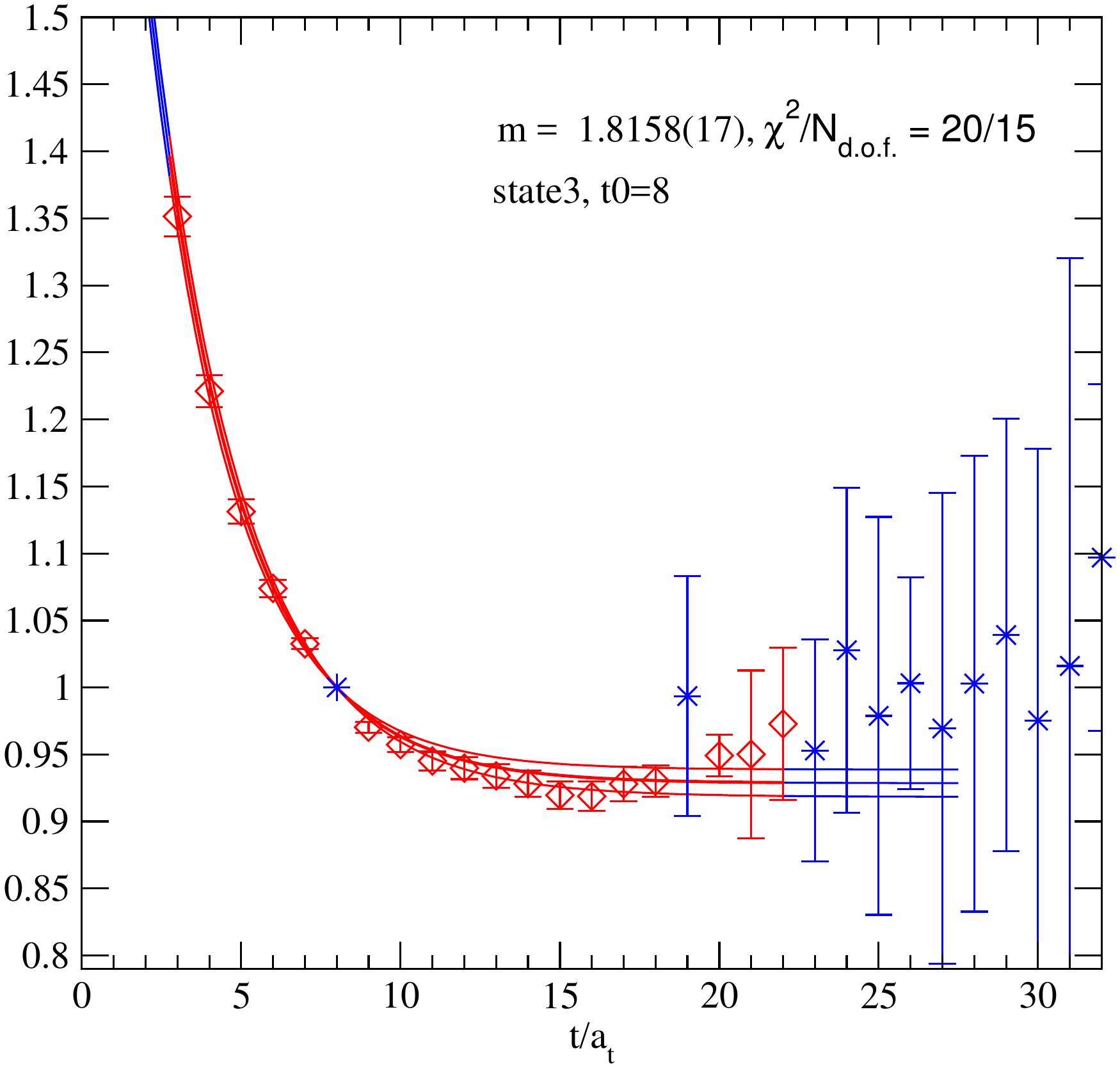}
    \caption{The principal correlator fits, for the ground, first, second and third excited states in the
      $T_1^{--}$ irrep in this work. The data points are
      $\lambda^n(t)\cdot e^{m_n(t-t_0)}$ with a reference timeslice $t_0=8$ as determined from fits to
      Eqn~\ref{eqn:princorrfits}. The extracted mass in lattice units together with the $\chi^2/N_{d.o.f.}$ are
      shown for each fit. The leading time dependence due to state $n$ has been divided out
      in each case, which results in a straight line at unity when a single exponential dominates the fit.}
  \label{fig:princorr}
  \end{center}
\end{figure}

\begin{figure}[h]
  \begin{center}
    \includegraphics[width=0.35\textwidth]{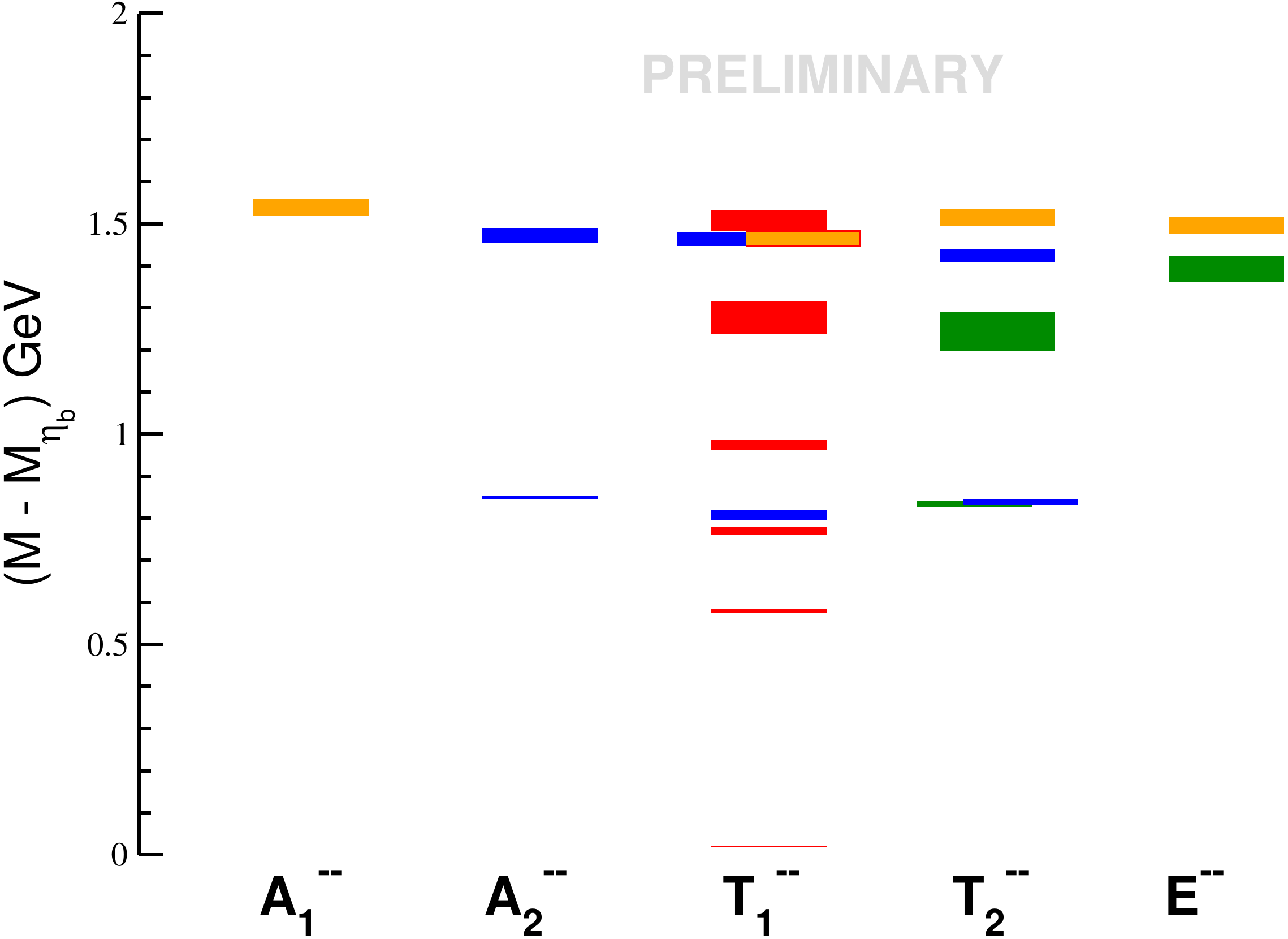}
    \includegraphics[width=0.35\textwidth]{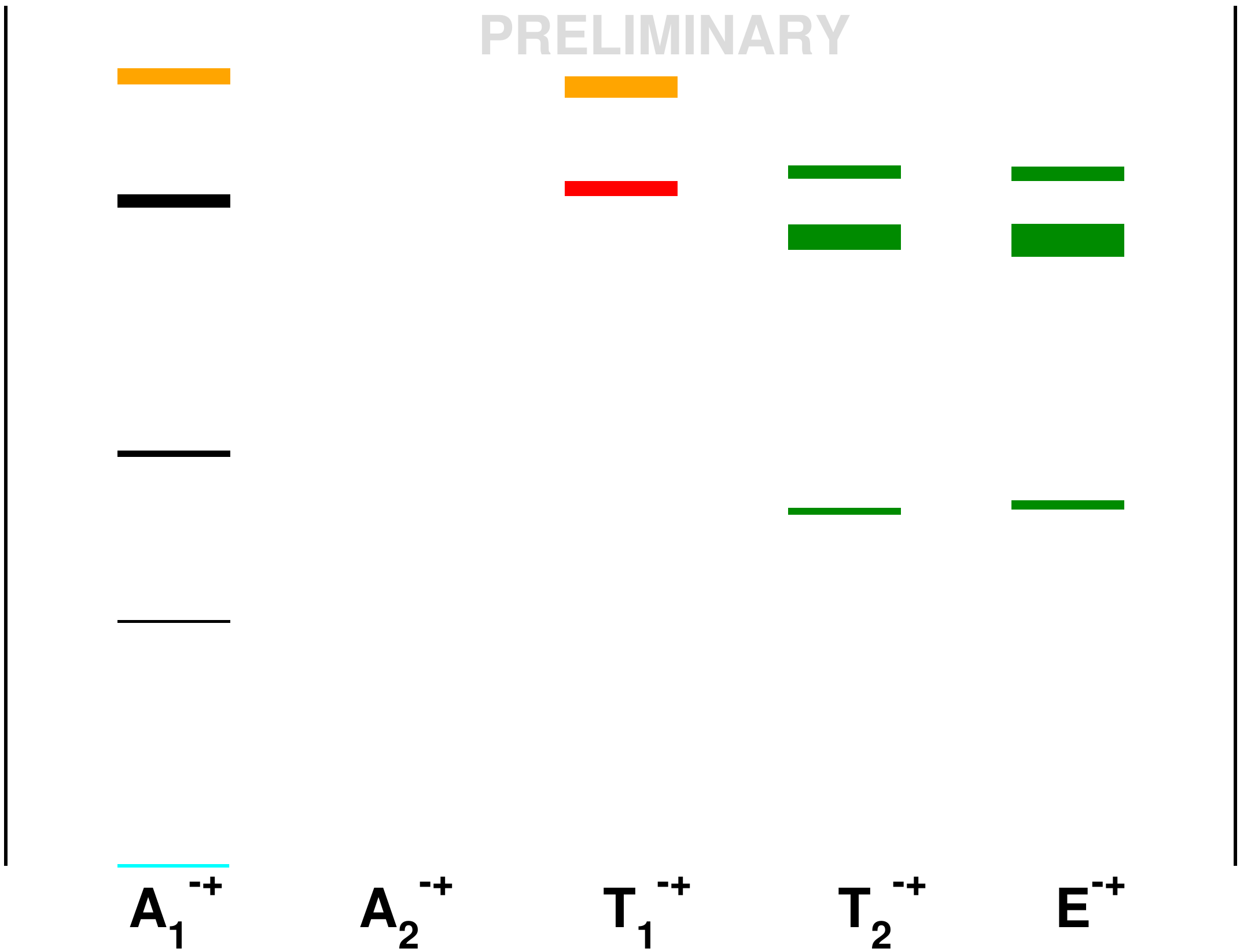}\\
    \includegraphics[width=0.35\textwidth]{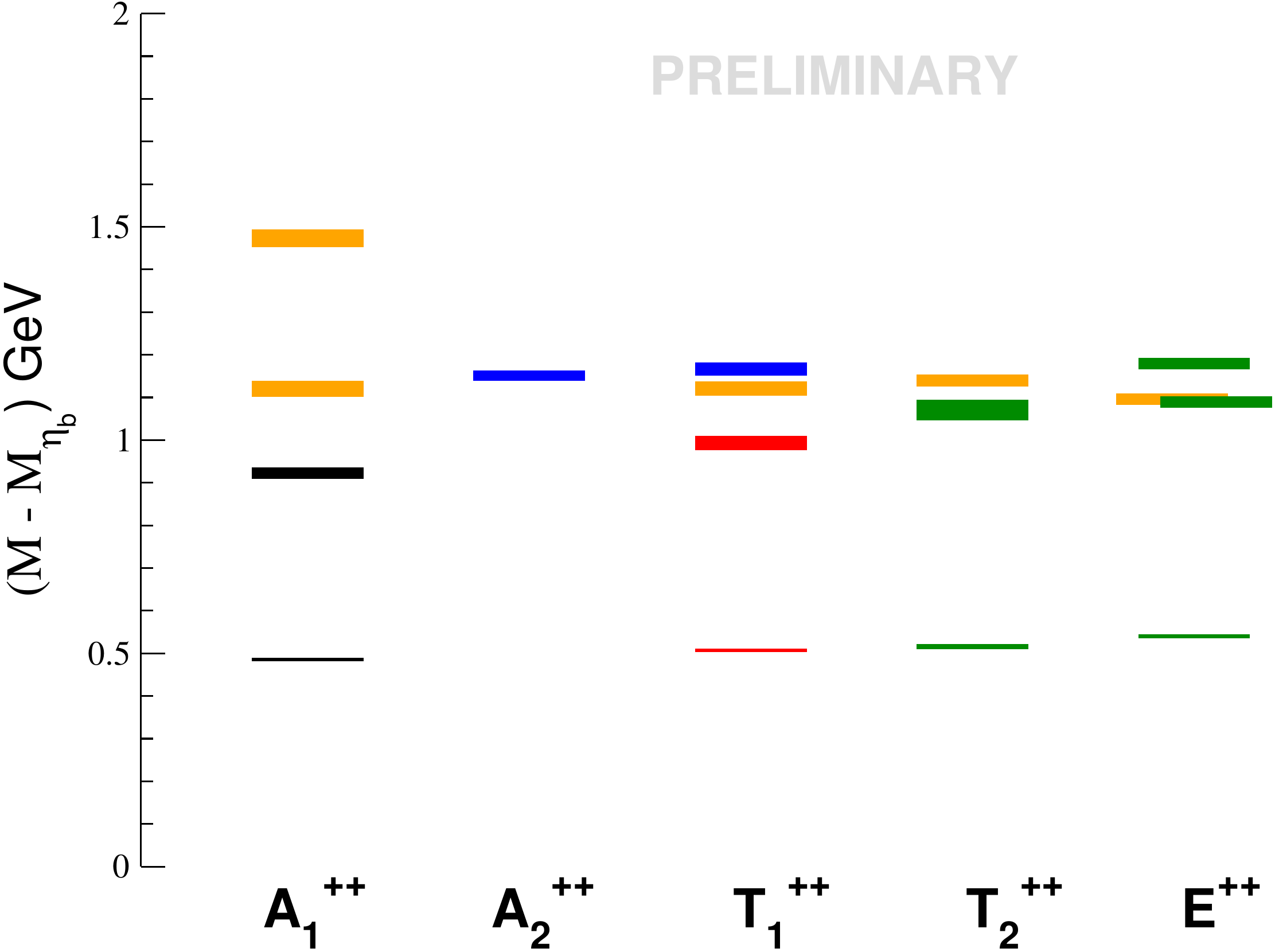}
    \includegraphics[width=0.35\textwidth]{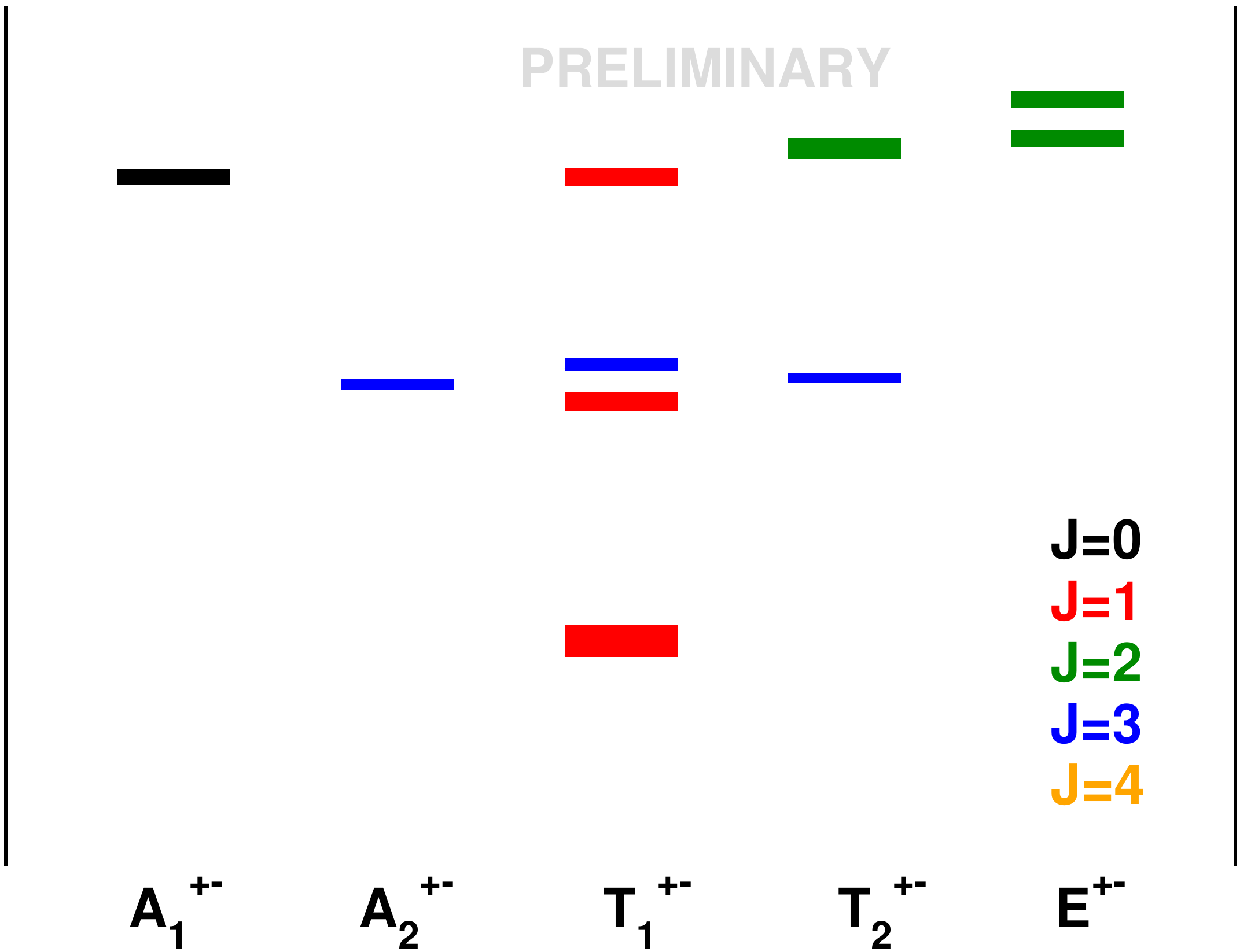}
  \caption{The bottomonium spectrum, labelled by lattice irrep $\Lambda^{PC}$, determined in this study. 
    The vertical height of the boxes represents the one-sigma statistical
    uncertainty about the mean determined from fits of principal correlators to
    Eqn~\ref{eqn:princorrfits}. Colours indicate the continuum spin.
    The spectrum is presented as mass splittings with the $\eta_b$ meson to reduce the uncertainty from tuning
    the heavy quark mass.}
  \label{fig:lattspec}
  \end{center}
\end{figure}

Local and non-local operators are constructed using distillation.
Following Refs.~\cite{Dudek:2009qf, Dudek:2010wm}
continuum operators are constructed with definite $J^{PC}$ and lattice versions using discrete
representations of the
gauge-covariant derivatives are then subduced to cubic group irreps, $\Lambda^{PC}$.  
The distillation method facilitates the efficient calculation of large operator bases,
including up to three derivatives as used in this study,
from which energies and overlaps are extracted by the
solution of a generalised eigenvalue problem at a suitable reference timeslice $t_0$. The energies are
determined from fits to the time-dependence of the eigenvalues (principal correlators) according to
\begin{equation}
  \lambda_n(t) = (1-A_n)e^{-m_n(t-t_0)}+ A_ne^{-m^\prime_n(t-t_0)},
  \label{eqn:princorrfits}
\end{equation}
with free parameters $A_n, m_n, m^\prime_n$. An example of the principal correlators in the
$T_1^{--}$ irrep from which energy levels are determined is shown in Figure~\ref{fig:princorr}.
The bottomonium spectrum, organised by lattice irrep is shown in Figure~\ref{fig:lattspec}.
The mass of the $\eta_b$ is subtracted to minimise uncertainty from the heavy quark tuning and
the energy levels are colour-coded according to the spin of the continuum operator which
dominates.
The pattern of states is broadly similar to that determined in an earlier study of the
charmonium spectrum~\cite{Liu:2012ze} including states with the exotic quantum numbers
$(0^{+-}, 1^{-+}, 2^{+-})$ found in $A_1^{+-}, T_1^{-+}$ and $(T_2^{+-},E^{+-})$,
and hybrid states which may have exotic or non-exotic quantum numbers. 
The $B_c$ spectrum is also determined yielding a ground state mass,
$m_{B_c}^{lattice}=6276 \pm 1 \;{\rm MeV}$, 
which can be compared with the experimental value~\cite{PhysRevD.98.030001}, $6274.9\pm 0.8$ MeV. A more extensive
calculation of the $B_c$ spectrum is underway. 
\subsection{The hybrid mesons in bottomonium}
\begin{figure}[h]
  \begin{center}
    \includegraphics[width=0.6\textwidth]{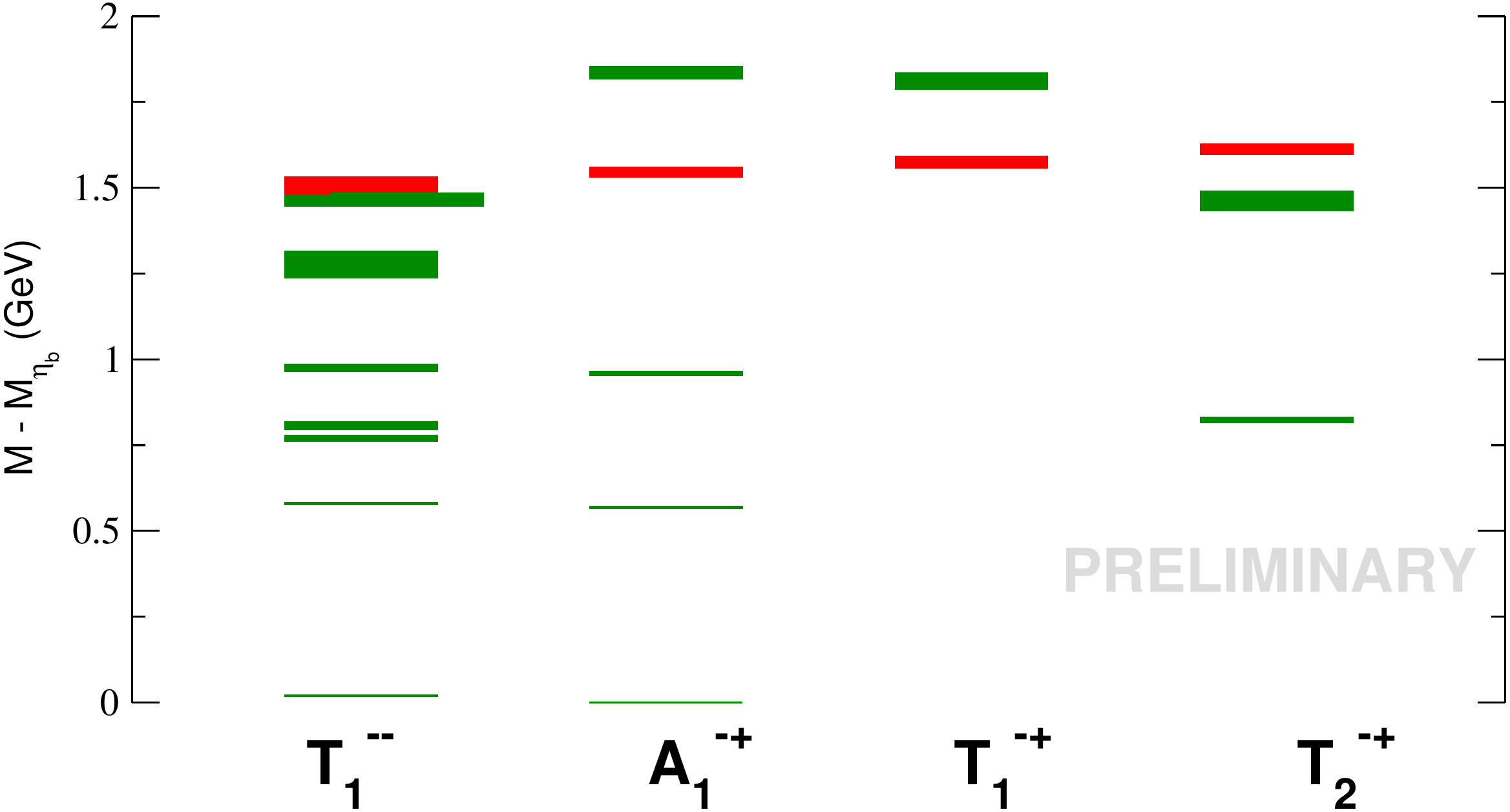}
    \caption{The lattice irreps in which candidate hybrid mesons are identified, by examining the
      operator-state overlaps.
      The energy levels are the same as those shown in Figure~\ref{fig:lattspec} with the hybrids shown in red.}
    \label{fig:hybrids}
    \end{center}
  \end{figure}

Following earlier studies in the light, open-charm and charmonium systems operators
with non-trivial gluonic content are included in the bases.
By analysing the operator overlaps candidate hybrid mesons are identified
when the state is dominated by a hybrid-like operator. In particular, three states are found which
do not seem to fit the pattern expected from quark models with
non-exotic quantum numbers, in the $T_1^{--}, A_1^{-+}$ and $(T_2,E)^{-+}$ irreps. These ``excess'' states have
a significantly larger overlap with operators proportional to the gluonic field strength tensor, $F^{\mu\nu}$ than the states in the same irreps which conform to a quark model picture.
An exotic hybrid meson is also identified in the the $T_1^{-+}$ irrep which is again
dominated by operators proportional to $F^{\mu\nu}$. 
The energy levels in these lattice irreps are shown in Figure~\ref{fig:hybrids} with the hybrid candidates shown in red.
The energy scale for the bottomonium hybrids (the splitting of the lowest-lying hybrid with the
lightest state in the system) is approximately $1.5$ GeV, similar to that
found in the light, heavy-light and charmonium meson and baryon sectors and suggesting similar
dynamics at play. Although still preliminary and not presented at this conference, the same
pattern of hybrids is also found in a analysis of the B-meson system, on the
same ensemble. These results also agree well with a recent determination of the spin-structure of
bottomonium hybrids~\cite{Brambilla:2018pyn} using non-relativistic
effective field theory and input from the lattice determination of the charmonium hybrids~\cite{Liu:2012ze}. 

\begin{figure}
  \begin{center}
    \includegraphics[width=0.6\textwidth]{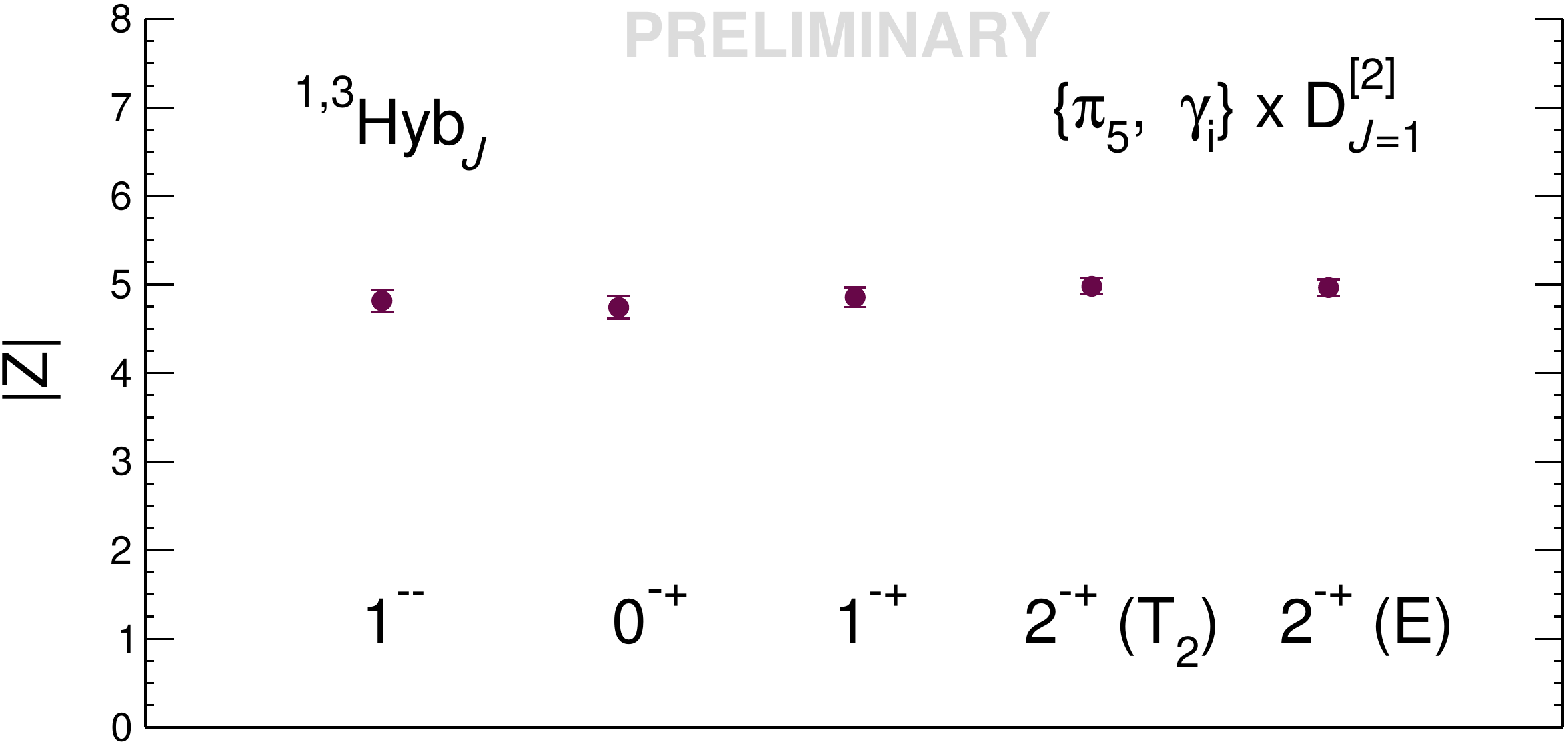}
  \end{center}
  \caption{The operator overlaps, $|Z|$, for the proposed
    lightest hybrid supermultiplet in bottomonium.}
  \label{fig:overlaps}
\end{figure}

Looking in more detail at the operator overlaps in the lattice irreps in
which hybrids have been identified and which were subduced from hybrid-like operators of the
form $\{\gamma_5,\gamma_i\}\times D_{J=1}^{[2]}$, a common absolute value of the overlap $|Z|$ is found,
encouraging the identification of a
lightest hybrid supermultiplet of bottomonium hybrids - corresponding to
those shown in Figure~\ref{fig:hybrids} - with $J^{PC} = (0,1,2)^{-+},1^{--}$. 
The overlaps determined in bottomonium channels are shown in Figure~\ref{fig:overlaps} and agree
very well with the similar study of charmonium, already mentioned~\cite{Liu:2012ze}.
\section{Summary}
An exploratory determination of the bottomonium spectrum is presented, including excited and exotic states up
to $J=4$. The calculation is performed on the Hadron Spectrum Collaboration's
$N_f=2+1$ dynamical anisotropic ensembles and fermions, including the bottom quark, are simulated with the same
relativistic action. The dispersion relations of pseudoscalar and vector heavy-heavy and heavy-light mesons show no
evidence of significant discretisation errors. 
An extensive spectrum of states is determined and includes the identification of a hybrid supermultiplet at an
energy scale similar to that found for hybrids in the light, heavy-light and charmonium sectors. Further work is ongoing,
including a determination of the $B$ and $B_c$ meson spectra.
\section*{Acknowledgements}
\noindent
SMR thanks colleagues in the Hadron Spectrum Collaboration and is grateful to
DAMTP, the University of Cambridge and the Technische Universit\"at M\"unchen
for hospitality while this work was ongoing. 

\end{document}